**Reaction operators for spin-selective chemical reactions of radical pairs**

J. A. Jones[a,b], Kiminori Maeda[b,c], P. J. Hore[d,*]

[a]Centre for Quantum Computation, Clarendon Laboratory, University of Oxford, Parks Road, Oxford OX1 3PU, UK

[b]Centre for Advanced ESR, University of Oxford, South Parks Road, Oxford OX1 3QR, UK

[c]Department of Chemistry, University of Oxford, Inorganic Chemistry Laboratory, South Parks Road, Oxford OX1 3QR, UK

[d]Department of Chemistry, University of Oxford, Physical and Theoretical Chemistry Laboratory, South Parks Road, Oxford OX1 3QZ, UK

[*]Corresponding author. Fax: +44 (0)1865 275410.
E-mail address: peter.hore@chem.ox.ac.uk (P.J. Hore).

**Abstract**

Spin-selective reactions of radical pairs have traditionally been modelled theoretically by adding phenomenological rate equations to the quantum mechanical equation of motion of the radical pair spin density matrix. More recently an alternative set of rate expressions, based on a quantum measurement approach, has been suggested. Here we show how these two reaction operators can be seen as limiting cases of a more general reaction scheme.

**1. Introduction**

Chemical reactions of free radicals are usually subject to conservation of electron spin angular momentum. For example, two radicals can normally only react to form diamagnetic products if there is some probability that their total electron spin is zero ($S = 0$, a singlet state). Conversely, in order to form a triplet-state product, the radical pair must have non-zero triplet character ($S = 1$). In general, singlet and triplet are not eigenstates of the spin Hamiltonian, so that a radical pair formed chemically in a singlet or triplet state will undergo coherent, quantum mechanical spin dynamics. Spin decoherence is often relatively slow so that weak internal and external magnetic interactions have time to modulate this quantum evolution. Such properties lead to a variety of interesting behaviour including a sensitivity of radical pair reaction rates and product yields to applied magnetic fields [1-4], chemically induced electron and nuclear spin polarizations [5] and magnetic isotope effects [6], phenomena that come under the umbrella of Spin Chemistry.

A proper treatment of spin-chemical effects requires the combination of a quantum mechanical description of the coherent spin dynamics, the kinetics of the spin-selective reactions, spin relaxation processes and relevant molecular motions. This is most conveniently performed using the stochastic Liouville equation

$$\frac{d\hat{\hat{\rho}}}{dt} = -\hat{\hat{V}}\hat{\rho}; \qquad \hat{\hat{V}} = i\hat{\hat{H}} + \hat{\hat{K}} + \hat{\hat{R}} + \hat{\hat{\Gamma}} \qquad (1)$$

where $\hat{\hat{H}}$ and $\hat{\hat{R}}$ are the Hamiltonian and relaxation superoperators, $\hat{\hat{K}}$ accounts for the reactivity and $\hat{\hat{\Gamma}}$ for the motion. This equation has the formal solution

$$\hat{\rho}(t) = \exp\left[-\hat{\hat{V}}t\right]\hat{\rho}(0). \qquad (2)$$

For many years the reaction term has usually been written using a phenomenological approach, often ascribed to Haberkorn [7], although the same form can be found in earlier works [8,9]. More recently an alternative form, based on quantum measurement theory, has been suggested [10] (see also [11-13]). Here we show how these two approaches can be viewed as limiting cases of a more general reaction scheme. (We do not consider here the ideas of Kominis [14-16], who suggests that an entirely different approach should be used.) For simplicity we ignore relaxation and motion in what follows, and write our density matrices in a minimal {S, T} basis, with S indicating a singlet state and T the $T_0$ ($m_S = 0$) triplet state. Initially we assume that reactions only occur from the singlet state, but this restriction is removed later.

## 2.    The conventional approach

The conventional description [7] of the dynamics of spin-sensing reactions is obtained by adding to the Liouville-von Neumann equation for the spin density matrix a rate equation for the disappearance of singlet terms at a rate $k_S$. This form of the equation can be justified in several ways, but perhaps the simplest is to begin with an unnormalised wavefunction

$$|\psi\rangle = c_S e^{-k_S t/2}|S\rangle + c_T|T\rangle \qquad (3)$$

where the *amplitude* of the singlet term disappears at a rate $k_S/2$, due to some process such as electron tunnelling to a product state. The corresponding density matrix is then

$$\hat{\rho} = |\psi\rangle\langle\psi| = \begin{pmatrix} c_S c_S^* e^{-k_S t} & c_S c_T^* e^{-k_S t/2} \\ c_T c_S^* e^{-k_S t/2} & c_T c_T^* \end{pmatrix} \qquad (4)$$

so the singlet *population* terms disappear at a rate $k_S$ as desired, while the off-diagonal coherent superposition terms only decay at a rate $k_S/2$. The equation of motion for the density matrix can be written as

$$\frac{d\hat{\rho}}{dt} = -i\left[\hat{H}, \hat{\rho}\right] - \frac{1}{2}k_S\left(\hat{\rho}\hat{Q}_S + \hat{Q}_S\hat{\rho}\right) \qquad (5)$$

where $\hat{Q}_S = |S\rangle\langle S|$ is the projection operator onto the singlet state. Rewriting the reaction terms in Liouville space in the {$|S\rangle\langle S|$, $|S\rangle\langle T|$, $|T\rangle\langle S|$, $|T\rangle\langle T|$} basis gives the form

$$\hat{\hat{K}}_{\mathrm{H}} = \frac{1}{2} k_{\mathrm{S}} \hat{\hat{Q}}_{\mathrm{S}}^{+} = \begin{pmatrix} k_{\mathrm{S}} & 0 & 0 & 0 \\ 0 & \tfrac{1}{2} k_{\mathrm{S}} & 0 & 0 \\ 0 & 0 & \tfrac{1}{2} k_{\mathrm{S}} & 0 \\ 0 & 0 & 0 & 0 \end{pmatrix} \qquad (6)$$

where the (anti)-commutator superoperators are given by

$$\hat{\hat{A}}^{\pm} = \hat{A} \otimes \hat{E} \pm \hat{E} \otimes \hat{A}^{\mathrm{T}} \qquad (7)$$

and $\hat{E}$ is the identity operator.

Clearly Equation (4) does not describe a proper density matrix, as its trace is not equal to one. This arises because the chemical reaction is included by the simple disappearance of singlet terms, without considering the reaction products, and can be overcome either by enlarging the description explicitly to include a term describing the reaction product P, or more simply by noting that the rate of appearance of product population is equal to the rate of disappearance of singlet reactant population. Following the former approach, the density matrix can be written in the enlarged {P, S, T} basis as

$$\begin{pmatrix} c_{\mathrm{S}} c_{\mathrm{S}}^{*} \left[1 - \mathrm{e}^{-k_{\mathrm{S}} t}\right] & 0 & 0 \\ 0 & c_{\mathrm{S}} c_{\mathrm{S}}^{*} \mathrm{e}^{-k_{\mathrm{S}} t} & c_{\mathrm{S}} c_{\mathrm{T}}^{*} \mathrm{e}^{-k_{\mathrm{S}} t/2} \\ 0 & c_{\mathrm{T}} c_{\mathrm{S}}^{*} \mathrm{e}^{-k_{\mathrm{S}} t/2} & c_{\mathrm{T}} c_{\mathrm{T}}^{*} \end{pmatrix} \qquad (8)$$

where the upper left corner corresponds to the singlet reaction products, and the lower right block describes the remaining radical pairs. The zeros in the top row and left hand column, connecting the radical pair with the reaction products, reflect the fact that coherent superpositions of different chemical species will rapidly decohere, and so can be safely neglected. This enlarged matrix is a proper density matrix, with trace equal to one.

More interestingly, the improper density matrix describing the remaining radical pairs always corresponds to a pure state; this is implicit in its derivation from a single wavefunction, and is easily confirmed by explicitly calculating the effective purity

$$\frac{\mathrm{Tr}(\hat{\rho}^{2})}{\left[\mathrm{Tr}(\hat{\rho})\right]^{2}} \qquad (9)$$

which is equivalent to calculating the purity after rescaling the density matrix to have unit trace. The significance of the slow decay of the off-diagonal terms is now clear: decay at a rate $k_{\mathrm{S}}/2$ is precisely the rate required to preserve the purity of the reactant state, with any more rapid decay leading to a mixed state for the remaining radical pairs. Note, however, that the enlarged density matrix, which explicitly includes the reaction products, is always mixed at $t > 0$ except in the trivial case that the radical pair begins in a pure triplet state.

## 3. Quantum measurements

An alternative approach to modelling spin-sensitive chemical reactions starts from the theory of quantum measurements [10]. We begin by considering a very simple case, which we subsequently develop into a model for spin sensing reactions. A projective (von Neumann) measurement on a quantum system is described by the process

$$\hat{\rho} \to \sum_m \hat{Q}_m \hat{\rho} \hat{Q}_m \qquad (10)$$

where the $\hat{Q}_m$ are projection operators corresponding to the possible measurement outcomes. Note that the density matrix is projected onto the measurement basis, causing coherent superpositions to collapse into mixed states. A chemical reaction which proceeds only for a singlet state could then be considered as a quantum measurement, collapsing the radical pair density matrix onto the measurement basis (here {S, T}), followed by the immediate conversion of the singlet component to reaction products, giving the final density matrix

$$c_S^* c_S |P\rangle\langle P| + c_T^* c_T |T\rangle\langle T|. \qquad (11)$$

It is rarely necessary to consider exactly *how* a von Neumann measurement occurs, but if necessary it can be considered as a limiting case of some physical process. For the case of singlet-sensing chemical reactions we note that Equation (8) reduces to Equation (11) in the limit of very long times or, equivalently, a very rapid reaction, so that the exponential terms tend to zero.

A pure von Neumann measurement of a radical pair would not be particularly interesting, as it would immediately remove all the off-diagonal terms which give rise to interesting behaviour. The quantum measurement model of spin-sensing reactions [10] is, however, more subtle. Suppose that during any short time d$t$ some fraction $k_S \, dt$ of radical pairs potentially undergo a chemical reaction, with the reaction succeeding if the spin system is in a singlet state and failing if it is in a triplet state. For the fraction which potentially reacts this reaction can be considered as a measurement of the spin state in the {S, T} basis as described above, in effect removing a fraction $k_S \, dt$ of the density matrix, and restoring only the triplet component. The remainder of the system, which does not even potentially react, is entirely unaffected, giving the equation of motion [10]

$$\frac{d\hat{\rho}}{dt} = -i\left[\hat{H}, \hat{\rho}\right] - k_S \hat{\rho} + k_S \hat{Q}_T \hat{\rho} \hat{Q}_T \qquad (12)$$

or in Liouville space

$$\hat{\hat{K}}_M = k_S \left(\hat{E} \otimes \hat{E} - \hat{Q}_T \otimes \hat{Q}_T\right) = \begin{pmatrix} k_S & 0 & 0 & 0 \\ 0 & k_S & 0 & 0 \\ 0 & 0 & k_S & 0 \\ 0 & 0 & 0 & 0 \end{pmatrix} \qquad (13)$$

Both the diagonal singlet population term and the off-diagonal superposition terms now decay at a rate $k_S$, and so the improper density matrix describing the remaining radical pairs initially becomes mixed, before finally tending to a pure triplet state at very long times, when all singlet components have been removed.

The additional dephasing of off-diagonal terms reflects an essential difference between the conventional and quantum measurement methods for treating the reaction. The conventional approach treats the spin sensing reaction as a continuous weak measurement, while the quantum measurement approach treats it as a probabilistic strong measurement. With a continuous weak measurement success indicates that the system is indeed in the singlet state, but failure indicates nothing, as a singlet state undergoing tunnelling could very well fail to show its singlet nature at any particular time. With probabilistic strong measurement, by contrast, failure indicates that the system is not in the singlet state, and thus must be in the triplet state, as the system would have shown success if it were in a singlet state. Non-reaction therefore constitutes a null measurement [17,18], and this null measurement provides the additional dephasing observed.

## 4. A general reaction scheme

As previously noted [10], the difference between the behaviour predicted by these two models is small, and can easily be masked by any additional sources of dephasing, such as spin relaxation. Nevertheless it is interesting to consider which of these models provides the more appropriate description of the reaction process. In fact, both models appear quite unrealistic, in that they include no dependence of the reaction rate on the molecular conformation. We now show how incorporating such dependencies into a model where the fundamental reaction process is based on electron tunnelling can lead to behaviour resembling either the conventional or the quantum measurement models.

We begin by considering charge recombination reactions of radical pairs in the solid state. Following simple Marcus theory, electron tunnelling between two radicals first requires the formation of a vibrationally excited state, corresponding to the crossing of the potential energy surfaces of the radical pair and the diamagnetic reaction products, followed by rapid electron transfer in an effectively static nuclear framework. Guided by these considerations, we treat a simple model in which the radical pair conformations are divided into two broad groups, reactive and unreactive.

We begin with Scheme 1: this includes two forms of the radical pair, $R_1$ and $R_2$ with identical spin Hamiltonians, which can undergo singlet-tunnelling to give the common reaction product P, at rates $k_{S1}$ and $k_{S2}$ respectively, and can also interconvert at rates $k_{12}$ and $k_{21}$.

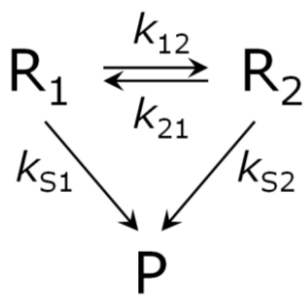

**Scheme 1**

The reaction operator in a stochastic Liouville equation for this general scheme is easily written using a $4 \oplus 4$-dimensional space, modelling the tunnelling using the conventional (i.e. Haberkorn) reaction operator and the interconversion process by classical chemical exchange terms to give an $8 \times 8$ matrix of the form

$$\hat{\hat{K}} = \begin{pmatrix} \mathbf{k_{S1}} + \mathbf{k_{12}} & -\mathbf{k_{21}} \\ -\mathbf{k_{12}} & \mathbf{k_{S2}} + \mathbf{k_{21}} \end{pmatrix} \quad (14)$$

where

$$\mathbf{k_{S1}} = \begin{pmatrix} k_{S1} & 0 & 0 & 0 \\ 0 & \tfrac{1}{2}k_{S1} & 0 & 0 \\ 0 & 0 & \tfrac{1}{2}k_{S1} & 0 \\ 0 & 0 & 0 & 0 \end{pmatrix}, \quad \mathbf{k_{12}} = \begin{pmatrix} k_{12} & 0 & 0 & 0 \\ 0 & k_{12} & 0 & 0 \\ 0 & 0 & k_{12} & 0 \\ 0 & 0 & 0 & k_{12} \end{pmatrix}, \quad (15)$$

and similarly for $\mathbf{k_{S2}}$ and $\mathbf{k_{21}}$. Modelling the interconversion classically implicitly assumes that superpositions of different molecular conformations will rapidly decohere, and can be safely neglected. The evolution under this operator can be determined from Equation (1) as usual, and the resulting density matrix then reduced to a $4 \times 4$ spin density matrix by tracing out the molecular conformation (as before [10], we initially evaluate the evolution under the kinetic operator, and subsequently add the Hamiltonian term back in). This can be simplified by noting that as the individual matrices in Equation (15) are diagonal the overall matrix, Equation (14), is block diagonal, containing four blocks of the form

$$\begin{pmatrix} \varepsilon k_{S1} + k_{12} & -k_{21} \\ -k_{12} & \varepsilon k_{S2} + k_{21} \end{pmatrix} \quad (16)$$

where $\varepsilon = 1$ for $\rho_{SS}$, $\varepsilon = \tfrac{1}{2}$ for $\rho_{ST}$ and $\rho_{TS}$, and $\varepsilon = 0$ for $\rho_{TT}$, each of which can be solved separately. The last case is trivial, as after tracing out the molecular conformation the exchange process has no effect on the $\rho_{TT}$ element, but the other three cases require explicit calculations.

The results are quite complex in the general case. Each term decays biexponentially: the two decay rates depend on all four rate constants in the kinetic scheme as well as the value of $\varepsilon$, and the amplitudes of the two terms depend also on the initial distribution of radical pairs between the two conformations $R_1$ and $R_2$. In three special cases, however, the behaviour is simple.

The first simple case is when $k_{S1} = k_{S2}$, so that the tunnelling rate is independent of conformation. After tracing out the molecular conformation, the reaction operator reduces to the conventional form, Equation (6), with $k_S = k_{S1} = k_{S2}$. This is easily rationalised: when both conformations are assumed to have the same tunnelling rate the exchange process can be ignored, and the reaction rate cannot depend on the exchange rates.

The second simple case is more interesting, and occurs when $k_{S2} \gg k_{21} \gg k_{12}$ and $k_{S1} = 0$ so that the major form $R_1$ is essentially unreactive while the minor component $R_2$ reacts extremely rapidly in comparison with the interconversion of $R_1$ and $R_2$. Denoting the density operators of $R_1$ and $R_2$, $\rho^{(1)}(t)$ and $\rho^{(2)}(t)$, respectively, one finds in this limit

$$\begin{aligned}\rho^{(1)}_{SS}(t) + \rho^{(2)}_{SS}(t) &= \rho^{(1)}_{SS}(0)e^{-k_{12}t} + \rho^{(2)}_{SS}(0)e^{-k_{S2}t} \\ \rho^{(1)}_{ST}(t) + \rho^{(2)}_{ST}(t) &= \rho^{(1)}_{ST}(0)e^{-k_{12}t} + \rho^{(2)}_{ST}(0)e^{-k_{S2}t/2}\end{aligned} \quad (17)$$

The terms depending on $\rho^{(2)}(0)$ can be safely dropped, as in the case $k_{21} \gg k_{12}$ the initial population of $R_2$ will be negligible. Thus the kinetic operator reduces to the quantum measurement form, Equation (13), with $k_S = k_{12}$. Again this result can be easily rationalised: rapid tunnelling aggressively dephases any states that can be dephased, effectively performing a projective measurement, and the overall reaction rate depends only on the rate at which the reactive state $R_2$ is formed, and not on the exact rate at which it then reacts.

Finally we consider the case when $k_{21} \gg k_{12} \gg k_{S2}$ and $k_{S1} = 0$ so that the major form $R_1$ is essentially unreactive as before, but the minor component now $R_2$ reacts only slowly in comparison with the interconversion between the two forms. In this case the kinetic operator reduces to the conventional form, Equation (6), with $k_S = k_{S2} \times k_{12} / k_{21}$. This third result also makes sense: the tunnelling is no longer fast enough to constitute a projective measurement, and so the kinetics are Haberkorn-like, with reaction only occurring from the small fraction $k_{12}/k_{21}$ of the molecules that are in a reactive conformation.

We therefore see that the conventional tunnelling model of spin-sensing reactions can lead to either the conventional (Haberkorn) reaction operator or to the quantum measurement operator, depending on the conformation-sensitivity of the reaction rate. If the reaction rate is only weakly dependent on conformation then the conventional result is recovered, but if the reaction rate is strongly dependent on conformation, with only a small fraction of possible conformations exhibiting any significant reactivity, then either Haberkorn-like or quantum measurement behaviour can be observed. In the latter case, the presence of a highly reactive conformation allows the tunnelling process to take on the characteristics of a strong projective measurement. In intermediate cases intermediate behaviour will be seen, with the off-diagonal elements exhibiting some additional dephasing, but not decaying quite as rapidly as the population terms.

For reactions in the solution state the situation is more complicated. Most radical pairs are effectively unreactive, as they are separated from one another by solvent molecules, but if the two radicals become trapped in the same solvent cage (a contact pair) then the inter-radical tunnelling reaction can be very rapid; thus the radical kinetics is dominated not by the reaction itself, but rather by the formation of contact pairs (the 'diffusion controlled' limit). This might seem equivalent to the situation in Scheme 1, but this is not the case, as the contact pair will have a quite different spin Hamiltonian from a separated radical pair, reflecting the strong distance dependence of the exchange interaction. Modulation of the exchange interaction by diffusion will act as a decoherence mechanism, further dephasing the off-diagonal terms [19-26]. This additional dephasing will outweigh any effects arising from the choice of reaction operator, rendering the choice between the two models largely moot.

## 5. Parallel singlet and triplet reactions

So far we have only considered reactions from the singlet state. The extension to radical pairs which react only from the triplet state is straightforward, as the behaviour of the S and $T_0$ states can simply be exchanged in the equations above. We assume for simplicity that the $T_{\pm 1}$ states do not interact with the other two states (as is the case in a strong applied magnetic field), so that their reaction kinetics is trivial.

The case of radical pairs which can react by *both* singlet and triplet pathways requires more thought. We begin by considering the predictions of the two models: if the reactions are independent with rates $k_S$ and $k_T$ then one can simply add operators for the two pathways, giving for the conventional approach

$$\hat{\hat{K}}_H = \frac{1}{2} k_S \hat{\hat{Q}}_S^+ + \frac{1}{2} k_T \hat{\hat{Q}}_T^+ = \begin{pmatrix} k_S & 0 & 0 & 0 \\ 0 & \tfrac{1}{2}(k_S + k_T) & 0 & 0 \\ 0 & 0 & \tfrac{1}{2}(k_S + k_T) & 0 \\ 0 & 0 & 0 & k_T \end{pmatrix} \tag{18}$$

and similarly for the quantum measurement approach

$$\hat{\hat{K}}_M = \begin{pmatrix} k_S & 0 & 0 & 0 \\ 0 & k_S + k_T & 0 & 0 \\ 0 & 0 & k_S + k_T & 0 \\ 0 & 0 & 0 & k_T \end{pmatrix} \tag{19}$$

An important special case occurs if $k_S = k_T$, when the conventional reaction operator is indistinguishable from that for a simple non-spin-sensing reaction, while the quantum measurement operator contains additional dephasing terms, which can be ascribed to the effects of null measurements [10,18] as before.

As before we now show how these two situations can be rationalised by considering the interplay between the tunnelling process and the molecular conformation. If the tunnelling rate is independent of conformation then it is clear that no interesting behaviour can arise, and the conventional description, Equation (18), applies. In the case $k_S = k_T$ the reaction is not in any sense spin-sensitive, and is entirely indistinguishable from any other chemical reaction which depletes the radical pair population. If the tunnelling rate does depend on conformation, however, then the possibility of more interesting behaviour does arise. We treat the interchange between molecular conformations as before, and consider two extreme cases: firstly, where there is a common reactive conformation, which can react by both the singlet and triplet pathways, and secondly where there are two quite separate intermediates for the two pathways.

In the first case the reaction can be described by an 8 × 8 matrix

$$\hat{\hat{K}} = \begin{pmatrix} \mathbf{k}_{12} & -\mathbf{k}_{21} \\ -\mathbf{k}_{12} & \mathbf{k}_{S2} + \mathbf{k}_{T2} + \mathbf{k}_{21} \end{pmatrix} \tag{20}$$

with

$$\mathbf{k}_{S2} = \begin{pmatrix} k_{S2} & 0 & 0 & 0 \\ 0 & \tfrac{1}{2}k_{S2} & 0 & 0 \\ 0 & 0 & \tfrac{1}{2}k_{S2} & 0 \\ 0 & 0 & 0 & 0 \end{pmatrix}, \quad \mathbf{k}_{T2} = \begin{pmatrix} 0 & 0 & 0 & 0 \\ 0 & \tfrac{1}{2}k_{T2} & 0 & 0 \\ 0 & 0 & \tfrac{1}{2}k_{T2} & 0 \\ 0 & 0 & 0 & k_{T2} \end{pmatrix} \qquad (21)$$

and when $k_{S2} = k_{T2}$ the matrix in Equation (20) is made up of four identical two-by-two blocks, so all four terms decay in the same manner. After making the approximations $k_{S2}, k_{T2} \gg k_{21} \gg k_{12}$ and ignoring the negligible contribution from the initial population of the minor component, this decay is found to be a single exponential at a rate $k_{12}$. Thus in this case the reaction kinetics is indistinguishable from a non-spin-sensing reaction with rate $k_{12}$.

In the second case it is necessary to use a 12 × 12 matrix. We assume that $R_1$ is entirely unreactive, while $R_2$ will react rapidly from the singlet state (but not from the triplet) and $R_3$ will react rapidly from the triplet state (but not from the singlet), so the kinetic operator is

$$\hat{\hat{K}} = \begin{pmatrix} \mathbf{k}_{12} + \mathbf{k}_{13} & -\mathbf{k}_{21} & -\mathbf{k}_{31} \\ -\mathbf{k}_{12} & \mathbf{k}_{S2} + \mathbf{k}_{21} & 0 \\ -\mathbf{k}_{13} & 0 & \mathbf{k}_{T3} + \mathbf{k}_{31} \end{pmatrix} \qquad (22)$$

where we have neglected direct exchange between the two reactive conformations. After tracing out the molecular conformation and making the usual approximations the quantum measurement operator, Equation (19), is recovered, with $k_S = k_{12}$ and $k_T = k_{13}$.

Thus, the behaviour of parallel singlet and triplet sensing reactions will depend on the fine details of the reaction process, and in particular on the interplay between the electron-tunnelling reaction and conformational changes. If the two reactions proceed at the same rate via the same intermediate then they cannot be considered in any practical sense spin-sensing. If, however, they proceed via different intermediate conformations then the two reactions can constitute quantum measurements even if they occur at the same rate.

## 6.    Conclusions

By considering two- and three-site models of radical pair reactivity, we have shown that a recombination operator with the same form as that derived from quantum measurement arguments can arise in multi-site models of radical pair reactions even though the elementary spin-selective reactions are described using the conventional Haberkorn approach. As before [10], the only difference between the two operators is that the quantum measurement approach predicts a doubling in the rate of singlet-triplet decoherence compared to the Haberkorn model. This additional dephasing is fundamentally indistinguishable from many other ubiquitous forms of dephasing. It would be very challenging, for example, to discriminate between dephasing arising from this source and that caused by stochastic fluctuations of spin interactions [11,19-26].

**Acknowledgements**

We thank the EMF Biological Research Trust for financial support. We gratefully acknowledge helpful discussions with participants at the workshop on 'Quantum measurement and chemical spin dynamics' held at the Lorentz Center in Leiden in March 2010.